\documentclass[a4paper,11pt]{article}

\usepackage{jheppub} 

\usepackage[T1]{fontenc} 
\usepackage[dvipsnames]{xcolor}
\usepackage[
 backend=biber,
 firstinits=true,
 sorting=none
]{biblatex}
\usepackage{lineno}
\usepackage{tikz}
\usetikzlibrary{calc}
\usepackage{siunitx}
\title{\boldmath Strategies for Beam-Induced Background Reduction at Muon Colliders}

\author[a]{D. Ally}
\author[a]{L. Carpenter}
\author[a]{T. Holmes}
\author[a,b]{L. Lee}
\author[a]{P. Wagenknecht}

\affiliation[a]{University of Tennessee, Knoxville, TN, USA}
\affiliation[b]{Harvard University, Cambridge, MA, USA}

\emailAdd{tholmes@utk.edu, llee@utk.edu}

\abstract{Future collider detectors at muon colliders will be bombarded by Beam-Induced Backgrounds (BIB) due to the in-flight muon decays from the beam line. These backgrounds can inhibit the ability of the detector and subsequent data analysis to successfully reconstruct collision products. We explore methods for geometrically reducing these effects for use in the readout, triggering, and data analysis of future experiments. Studies are performed for a collision energy of 1.5~TeV, and a detector with a tungsten nozzle designed to block the majority of the BIB. In this context, detector strategies are explored to further reduce the BIB, with a focus on the innermost layers of the tracker where its density is highest. In addition, a conceptual design of a calorimeter built to reject BIB is presented.

}

\begin{document} 
\maketitle
\flushbottom

\setlength\parindent{0pt}
\setlength{\parskip}{0.5em}

\newcommand{\pt}{$p_{\mathrm{T}}~$}
\newcommand{\dz}{$d_{0}~$}

\section{Introduction}

Muon colliders are a promising option for future colliders: they are arguably the most cost-, space-, and energy-efficient way to explore the energy frontier \cite{Long:2020wfp}. From the perspective of detector performance, the largest blocker for a successful muon collider is the Beam-Induced Background (BIB) \cite{muc-perf-2020, Mokhov:2011zzd, Bartosik:2019dzq}. The metastable muon decays in flight, producing a halo of secondary particles that travel along with the beam. As in previous works \cite{Bartosik:2019a9}, a detector design employing a tungsten nozzle to block this BIB is used. 

Even with this shield in place, the remaining BIB drowns out the signals of particles from the collision. While the BIB extends throughout the tracker and into the calorimeter, we chose to first address this problem where it is the most intense, in the detector elements closest to the beam. With the exception of Section~\ref{sec:calo}, this paper explores strategies to reduce BIB at the hit level in the innermost layers of the tracker, to reduce this overwhelming background before it becomes a combinatorial blocker for tracking algorithms.

\section{Simulated Samples}

Simulated samples were provided by the International Muon Collider Collaboration (IMCC) \cite{imcc} or produced using its software as made available for the Snowmass process. The BIB background overlay was provided by the larger community (originally produced in MARS15 by N. Mokhov~\cite{osti_1282121}). Di-Higgs production is used as an example signal process in these studies, and all Higgs bosons are assigned Standard Model couplings. These events were produced privately, generated in \textsc{MadGraph5\_aMC@NLO}~\cite{Alwall:2014hca} interfaced with \textsc{Pythia8}~\cite{pythia2015}, and the provided \texttt{Marlin} framework~\cite{Gaede:2003ip}. A beamspot spread of 1 cm was used. The results presented here were produced using energy depositions from particle interactions as simulated by \textsc{Geant4}~\cite{AGOSTINELLI2003250}. Each simulated energy deposition will be referred to as a \emph{sim hit}.

The detector geometry assumed in the simulation is a modified CLIC detector design, \texttt{CLIC\_o3\_v14\_mod4}, and the majority of these results focus on the vertex detector, the innermost silicon tracking system, with a tungsten nozzle added on either side of the detector to block incoming BIB \cite{Bartosik:2019a9}. 

\section{BIB Discrimination Strategy in Trackers}

The primary source of BIB in the detector is not direct muon decay products, but the low-energy particles resulting from those decay products interacting with the nozzle. To travel from the nozzle to the vertex detector, these particles must have a shallow angle with respect to the beam line, and they can arrive in a large time window.

In this study, the target is to retain sensitivity to collision product particles originating at the interaction point (IP) which could feasibly be tracked, here defined as those with momentum greater than 1 GeV. In the simulated collision events, any vertex detector hit associated with a charged particle matching these requirements is classified as signal. All hits from BIB contributions are considered background, regardless of particle momentum. A fiducial region in time is defined such that only sim hits within the range $[-250,+300]$~ps with respect to the expected time of arrival assuming $\beta=1$ are included.

Even after this fiducial cut on the BIB, it is overwhelmingly large compared to signal. In the samples used here, a typical event contained approximately 600,000 hits from BIB in the vertex detector, but only about 250 signal hits.

These two populations have strikingly different features at the track level, but discrimination is needed before this stage to make tracking feasible. In this study, we explore two strategies for hit-level BIB reduction: one using precision timing, and one using double tracker layers to make angular measurements from hit pairs. 

\subsection{Timing Measurements}

\begin{figure}[tb]
    \centering
    \includegraphics[width=0.65\textwidth]{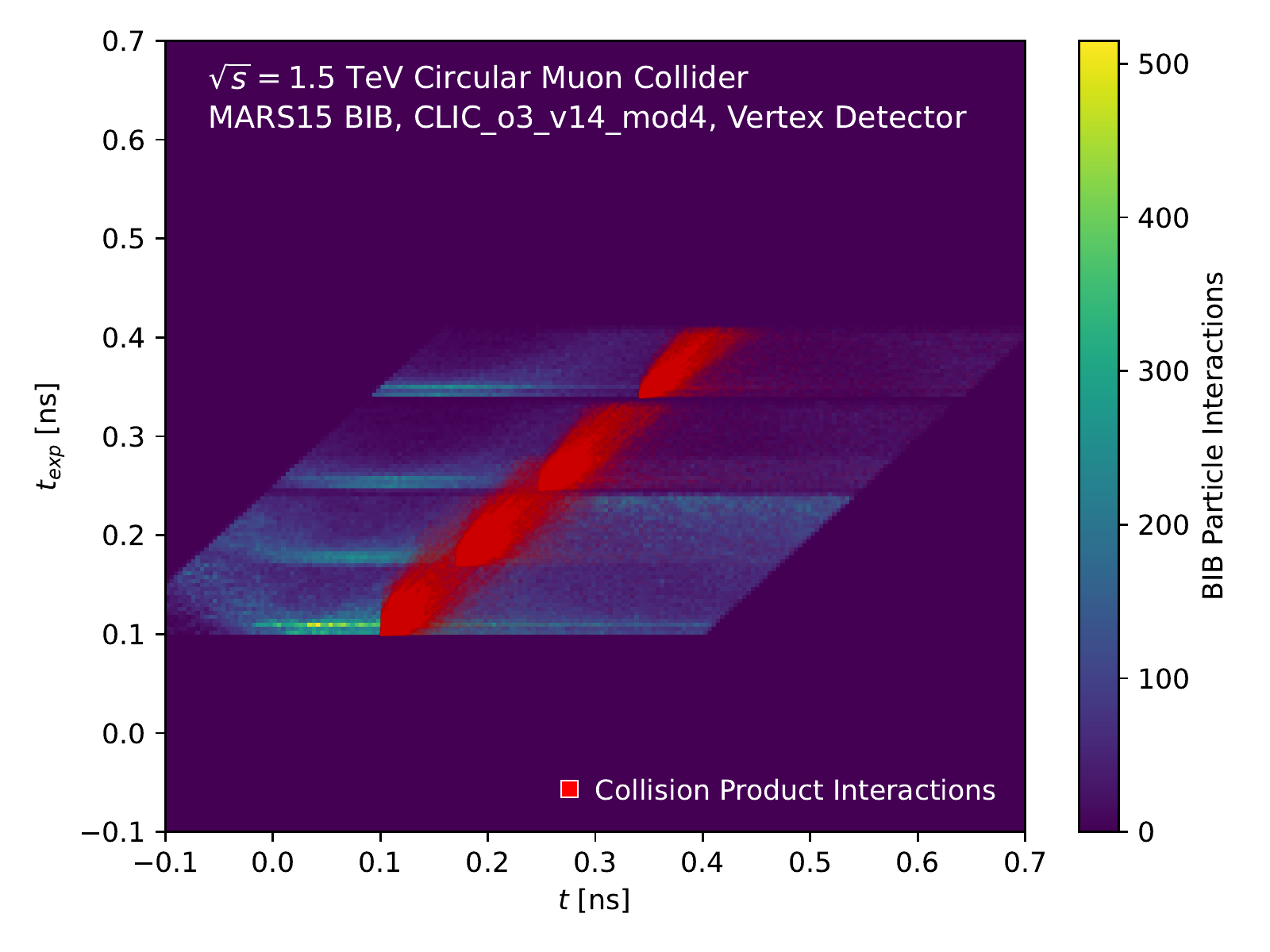}
    \includegraphics[width=0.45\textwidth]{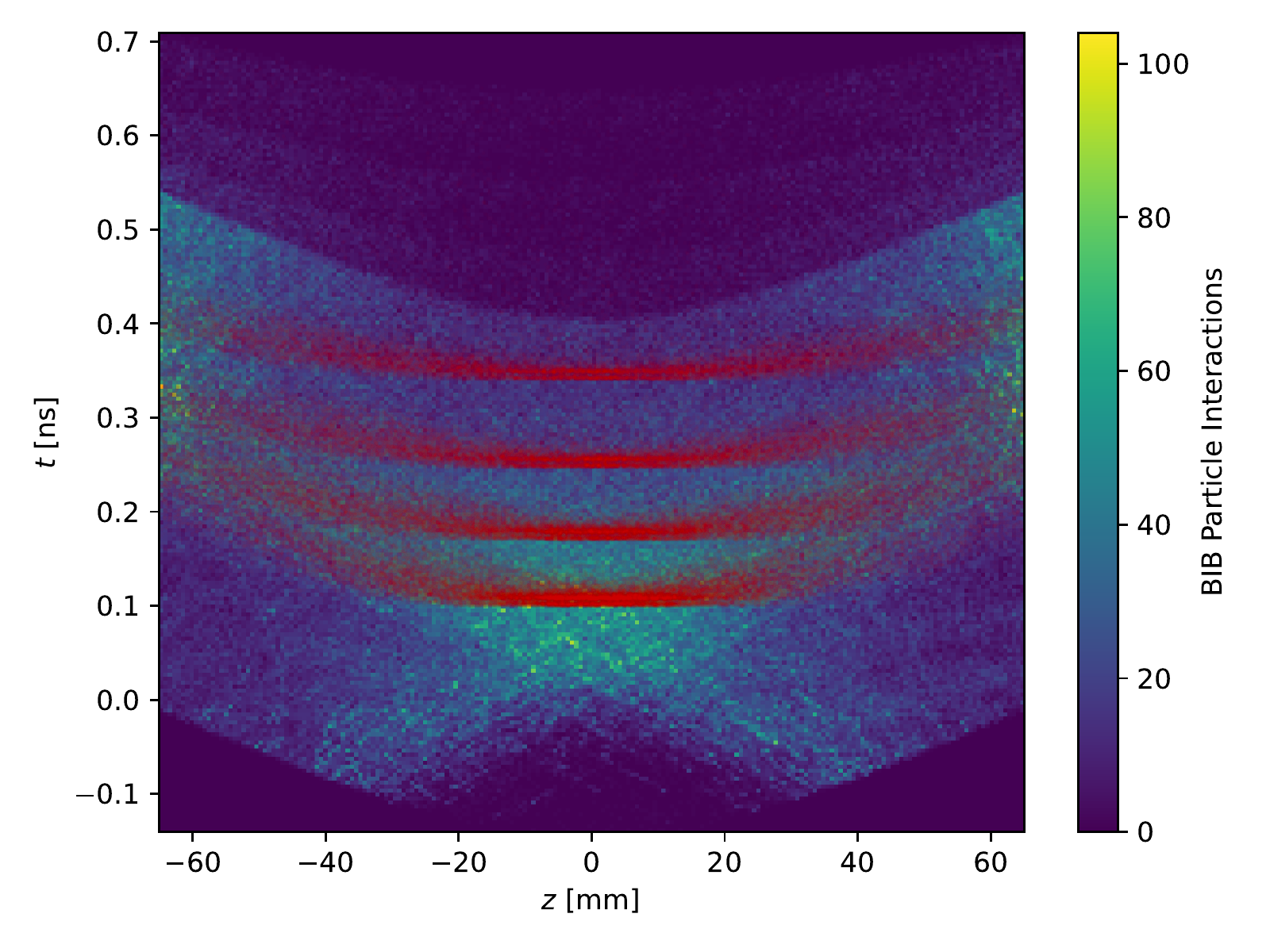}
    \includegraphics[width=0.45\textwidth]{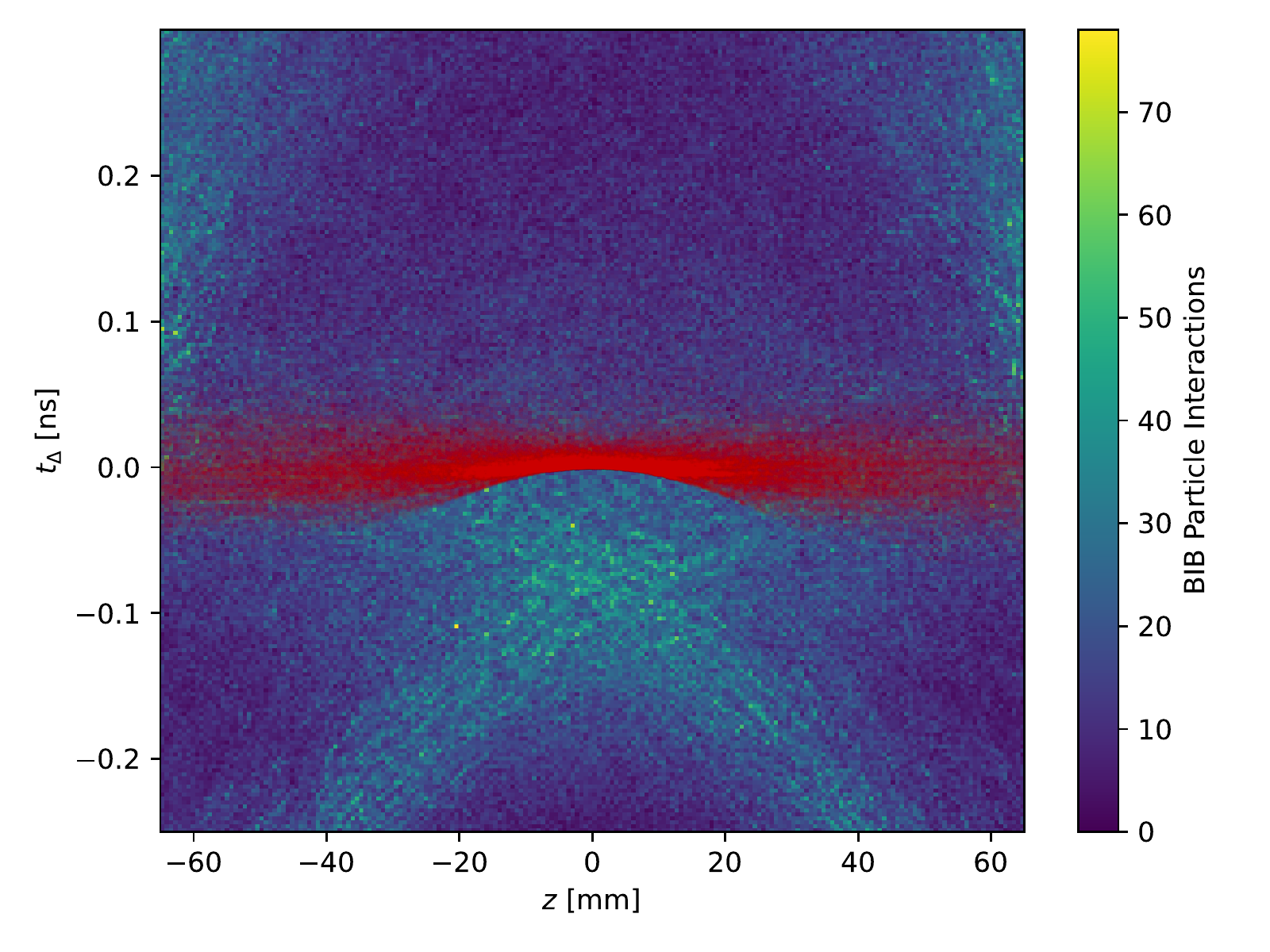}
    \caption{[Top] The correlation between the hit time $t$ and the expected hit time $t_{\rm exp}$ is shown. The red contribution from the collision products shows a strong correlation, while the color scale shows the structure due to BIB contributions. The measured time $t$ [Bottom Left] and relative time $t_{\Delta}$ [Bottom Right] as a function of longitudinal position $z$ are also shown.
    }
    \label{fig:time}
\end{figure}

Precise timing measurements can serve to reject energy depositions due to BIB, particularly in the tracker. Each detector element's position has a particular expected time of arrival $t_{exp}$ for particles with relativistic $\beta=1$. As shown in Figure~\ref{fig:time}, for particles from the collision, the actual time of arrival $t$ is highly correlated with $t_{exp}$. The energy depositions due to BIB do not show such a correlation and often arrive to the sensor earlier than expected (and earlier than allowed by the speed of light) if originating at the IP. The corrected time 

\begin{align}
    t_\Delta = t-t_{exp}
\end{align}

\noindent should strongly peak at zero for particles from the IP, while the BIB has a characteristic timing distribution similar to the distribution seen for BIB at the LHC. These behaviors can be seen in Figure~\ref{fig:time}. Selecting for small values of $t_\Delta$ will retain the interactions due to collision products while removing a large amount of the BIB.

While strict requirements on $t_\Delta$ may reduce BIB contributions, they may also reduce the acceptance for low-$\beta$ particles and may negatively affect searches for heavy BSM long-lived particles either produced with a $\beta<1$ or that have non-trivial flight paths from intermediate displaced decays~\cite{Lee_2019}.

\subsection{Angular Measurements}

\begin{figure}[tb]
    \centering
    \includegraphics[width=0.65\textwidth]{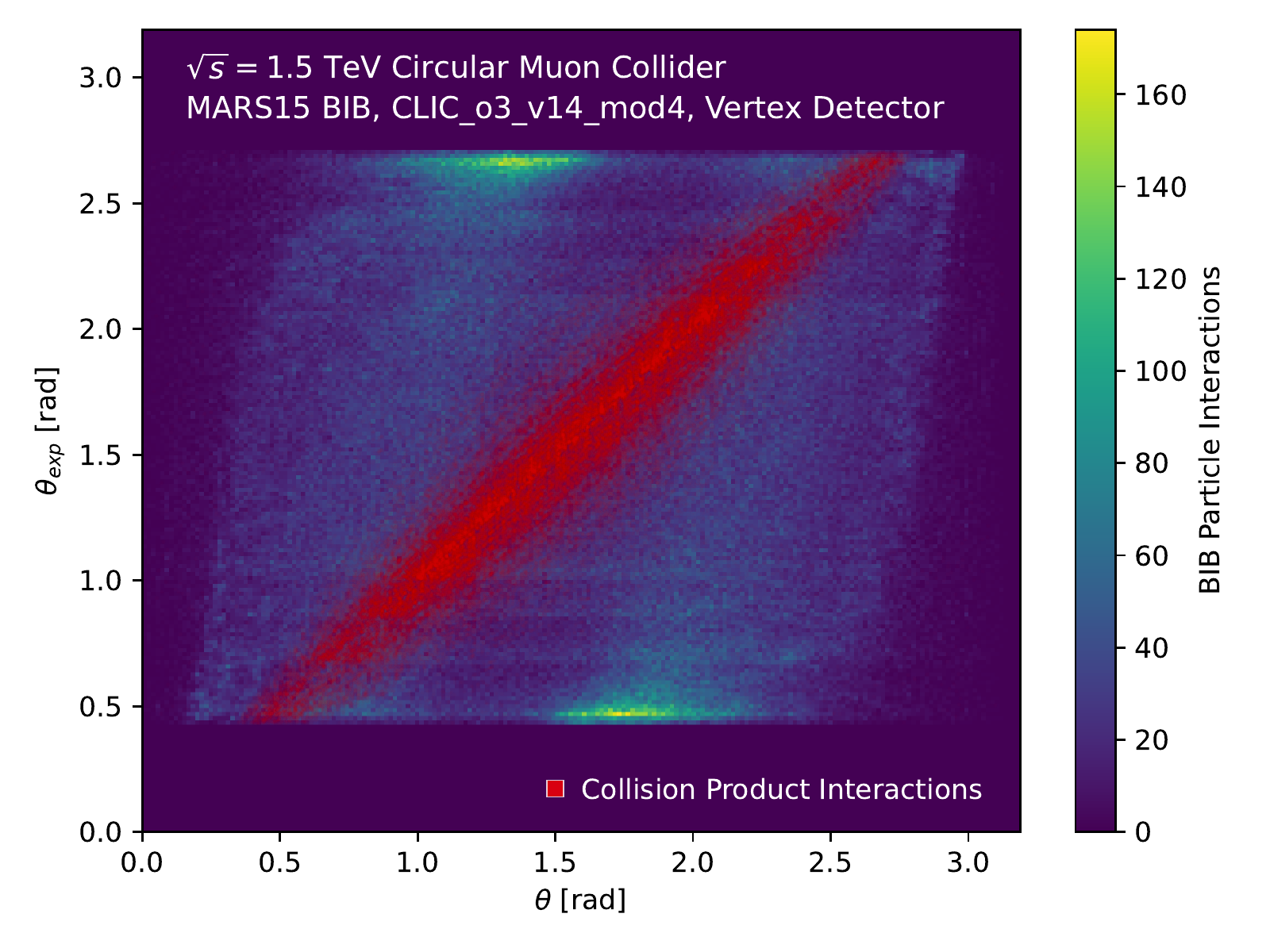}
    \includegraphics[width=0.45\textwidth]{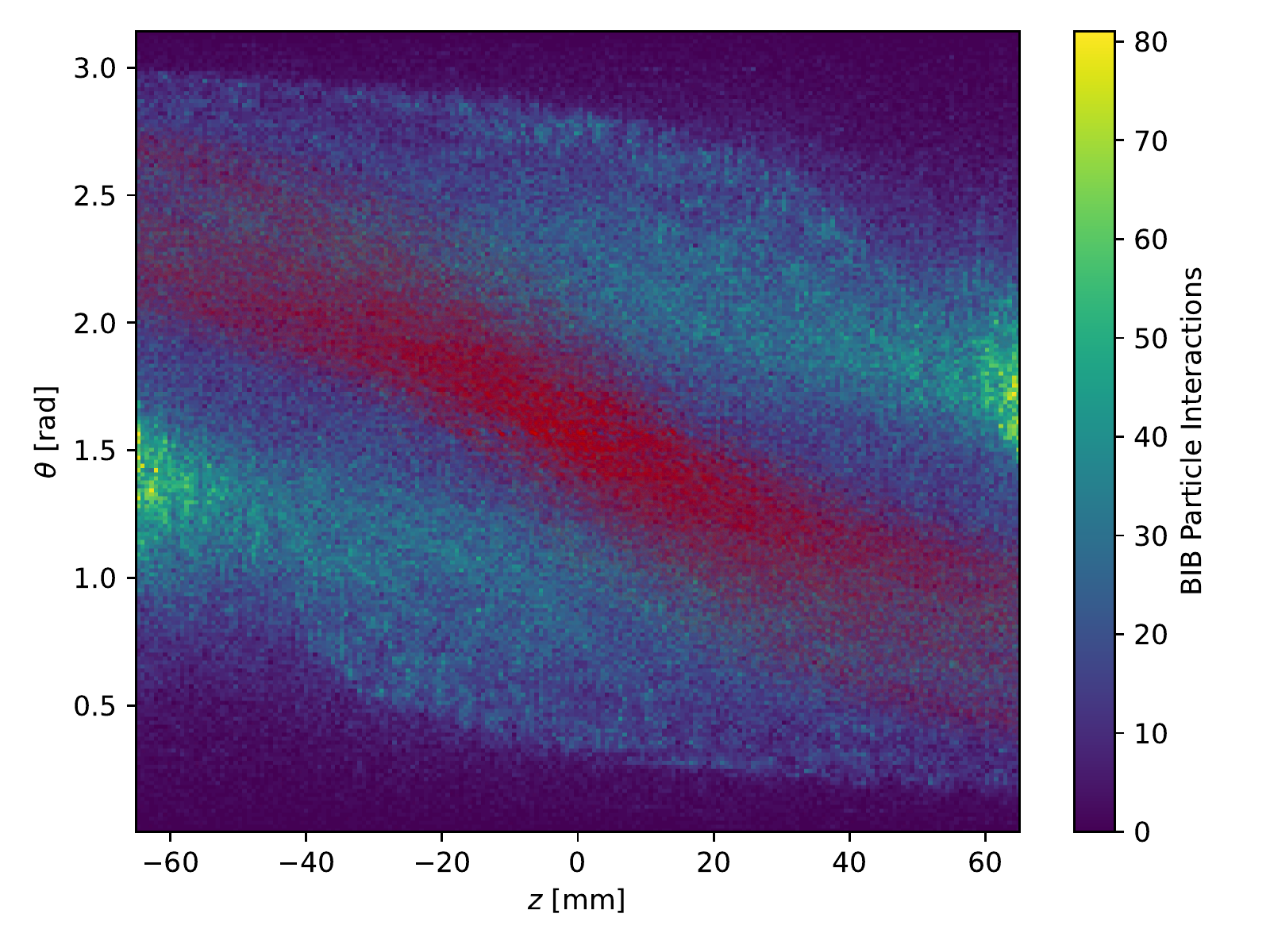}
    \includegraphics[width=0.45\textwidth]{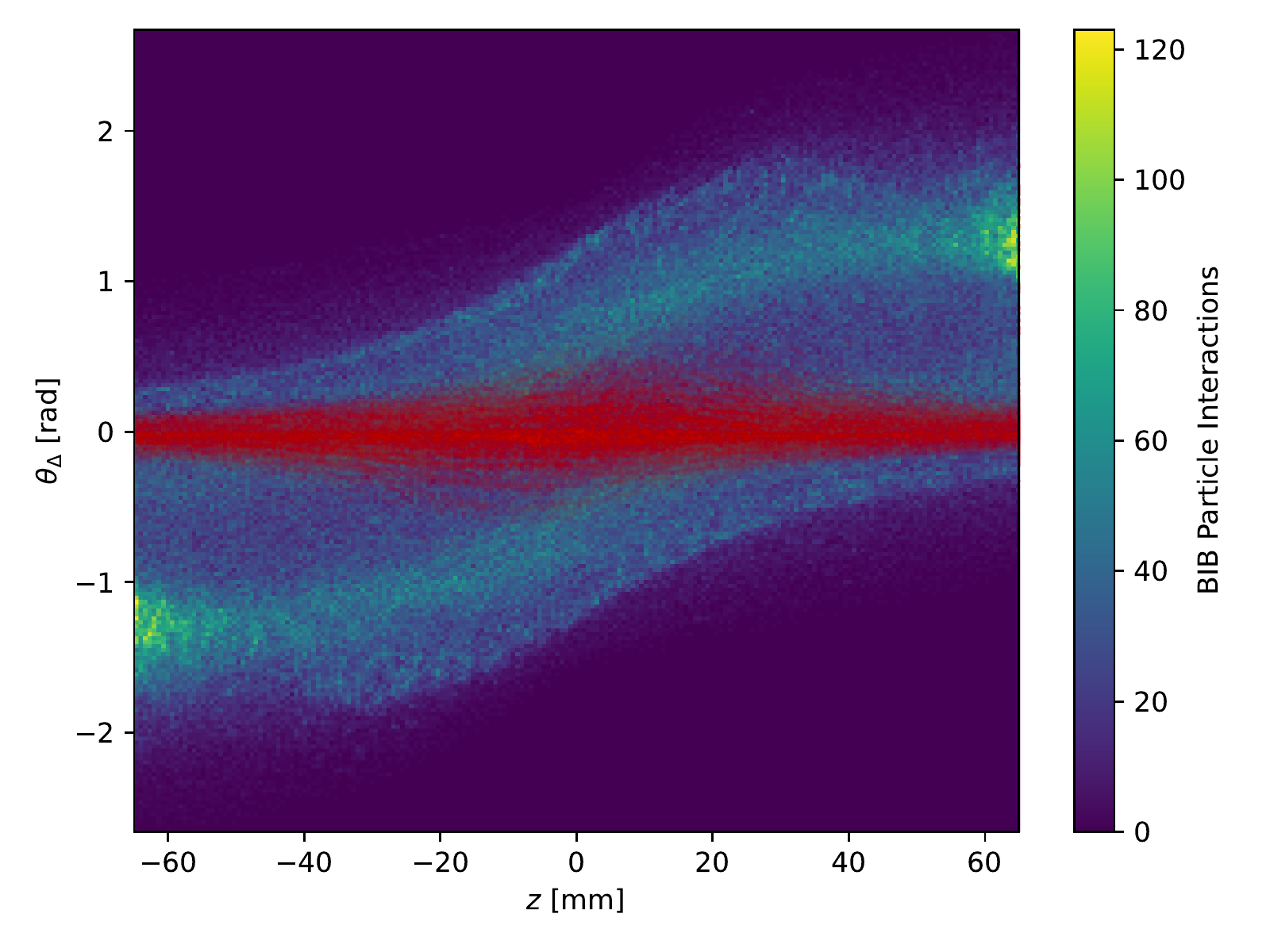}
    \caption{[Top] The correlation between the incident polar angle $\theta$ of the particle creating a hit and the expected angle $\theta_{\rm exp}$ is shown. The contribution from the collision products (red) shows a strong correlation, while the color scale shows the structure due to BIB contributions. The measured angle $\theta$ [Bottom Left] and relative angle $\theta_{\Delta}$ [Bottom Right] as a function of longitudinal position $z$ are also shown.
    }
    \label{fig:angleDists}
\end{figure}

As in the case for timing, particles emerging from the IP have an expected incident angle $\theta$, while interactions from BIB may arrive at the sensors at other angles. Figure~\ref{fig:angleDists} shows that particles from the IP have a $\theta$ value well correlated with the expected $\theta_{exp}$ that will vary around the volume of the detector. BIB interactions are not well correlated in this plane, but in fact show a distinctive structure, with each beam producing BIB hits with incidence angles biased towards the original beam direction.

Once again, the correlation for particles from the IP can be exploited by a corrected $\theta$ variable,

\begin{align}
    \theta_\Delta = \theta-\theta_{exp},
\end{align}

\noindent for which the collision products have a value close to zero, while BIB contributions are found at other preferred angles that vary across the detector.

In order to take advantage of this feature, the tracker would need to provide a high-resolution $\theta$ measurement for incoming particles. This could be achieved via doublet layers of silicon giving a stub-based $\theta$ measurement, or via some other technology. In the case of doublet layers, the angular resolution depends on the pitch, the distance between layers, and the incident angle. For a 50 $\mu$m pitch with a 2 mm distance between layers, the maximum angular resolution is about 0.025 radians.

As in the case of the timing measurements above, strict requirements on $\theta_\Delta$ may also reduce the acceptance for BSM long-lived particles producing non-trivial flight paths from intermediate displaced decays~\cite{Lee_2019}.

\subsection{Combined Discrimination}

\begin{figure}[tb]
    \centering
    \includegraphics[width=0.8\textwidth]{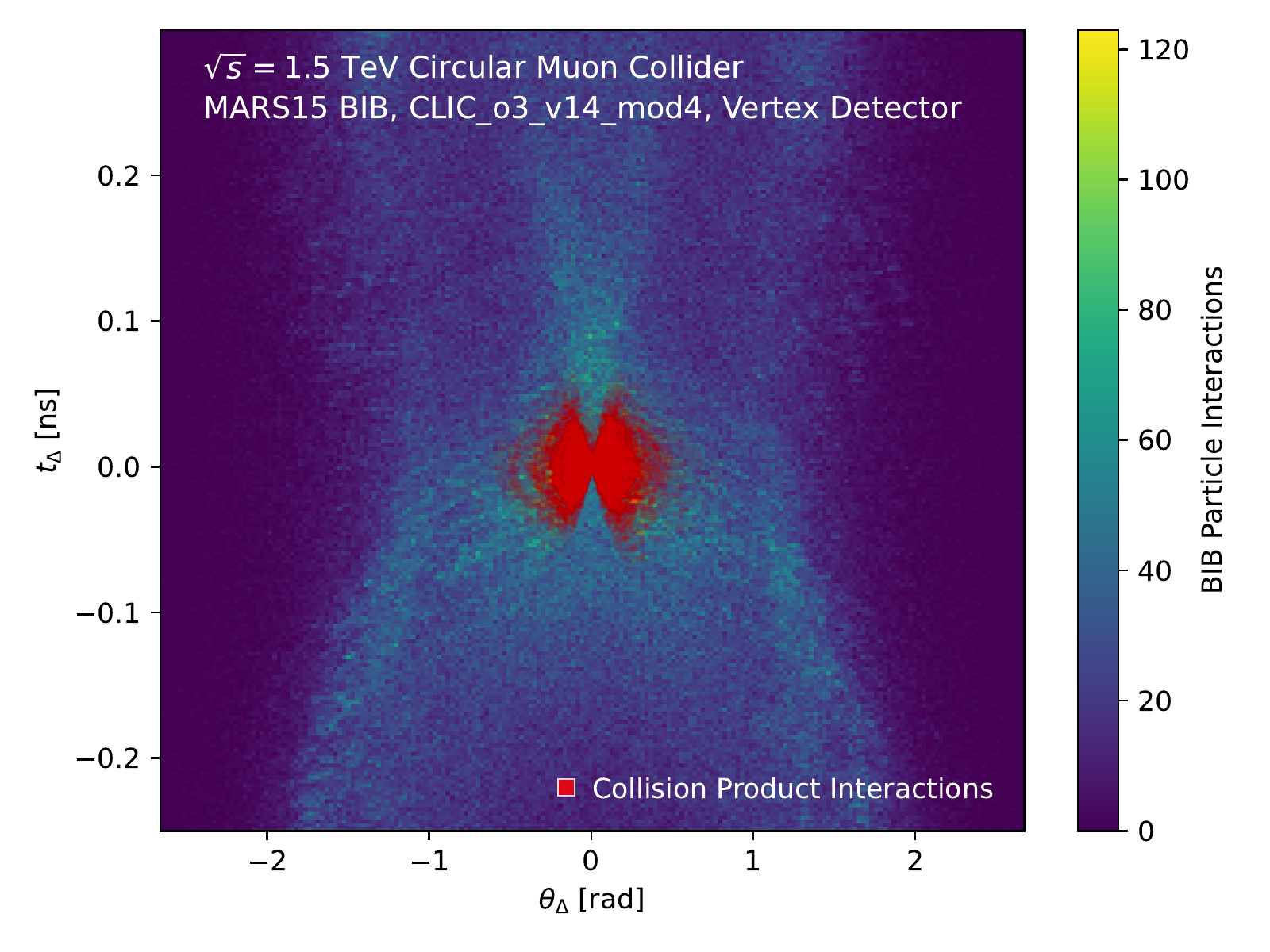}
    \caption{The two-dimensional distribution of $t_\Delta$ vs. $\theta_\Delta$ is shown. The contribution from the collision products (red) is largely localized at the origin with a tail extending to positive $t_\Delta$, while the color scale shows the structure due to BIB contributions.
    }
    \label{fig:tDeltaVsThetaDelta}
\end{figure}

For mitigating BIB in the inner tracking detector, we explore simultaneous requirements on $t_\Delta$ and $\theta_\Delta$. This space, shown in Figure~\ref{fig:tDeltaVsThetaDelta}, has contributions from collision products localized at (0 rad, 0 ns) with a tail extending up to positive $t_\Delta$ values. The BIB contributions, while structured, extend to large values of both $t_\Delta$ and $\theta_\Delta$. Selecting increasingly small rectangles in this space can serve to reject the BIB contributions while retaining particles emerging from the IP.

\begin{figure}[tb]
    \centering
    \includegraphics[width=0.8\textwidth]{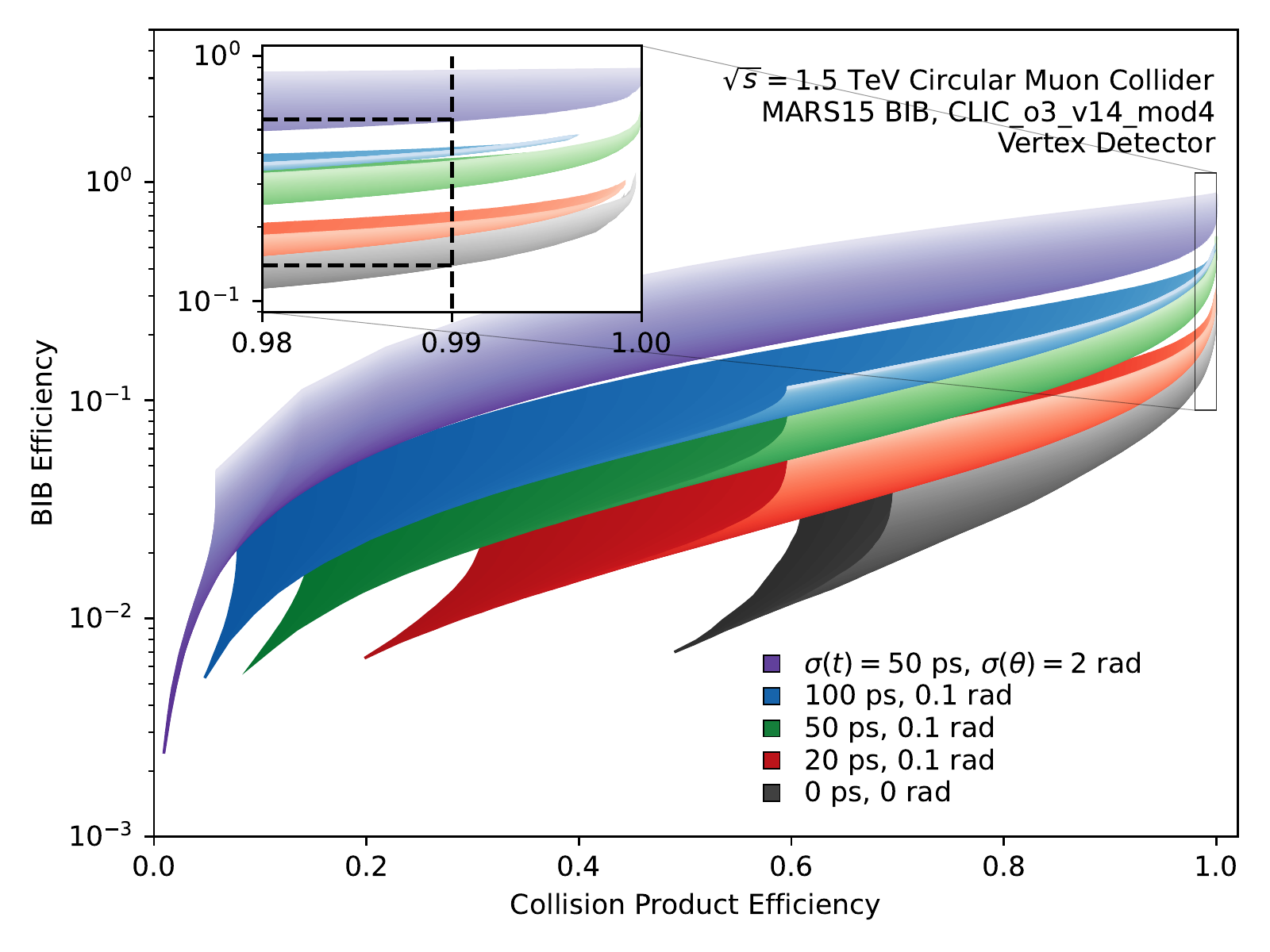}
    \caption{ROC surfaces are shown for a few example resolution values $\sigma(t)$ and $\sigma(\theta)$. Each resolution configuration is shown in a different hue. The darkness of the color represents the $t_\Delta$ requirement, while each $\theta_\Delta$ requirement (shown in the same hue and darkness) forms a traditional, one-dimensional ROC curve. Different hue ribbons represent different detector capabilities, while the surface of the ribbon represents the potential performance of the discriminators described in the text. Efficiencies are shown after a fiducial $t_\Delta$ selection of [-250, +300] ps.}
    \label{fig:ROC}
\end{figure}

\begin{figure}[tb]
    \centering
    \includegraphics[width=0.45\textwidth]{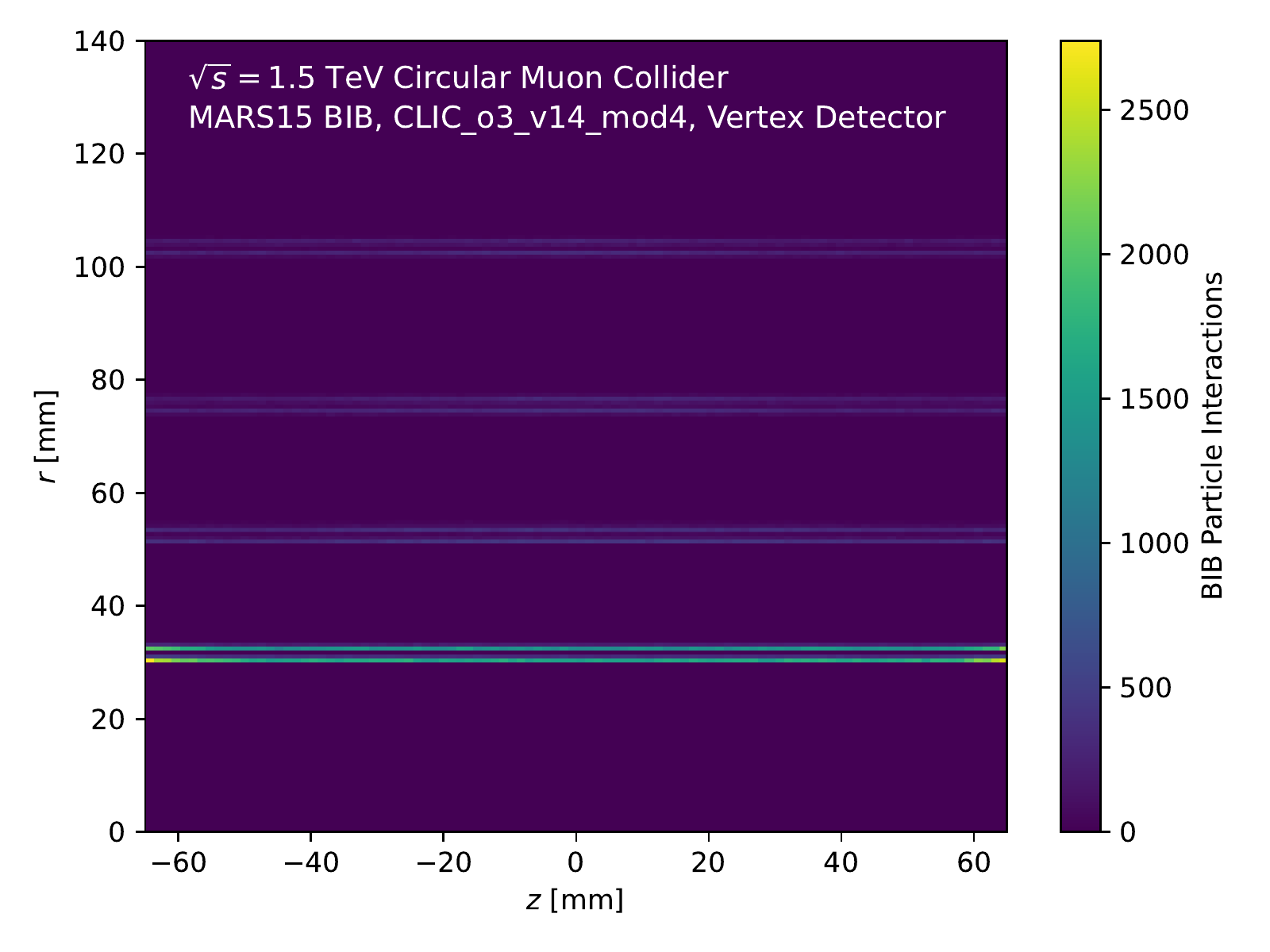}
    \includegraphics[width=0.45\textwidth]{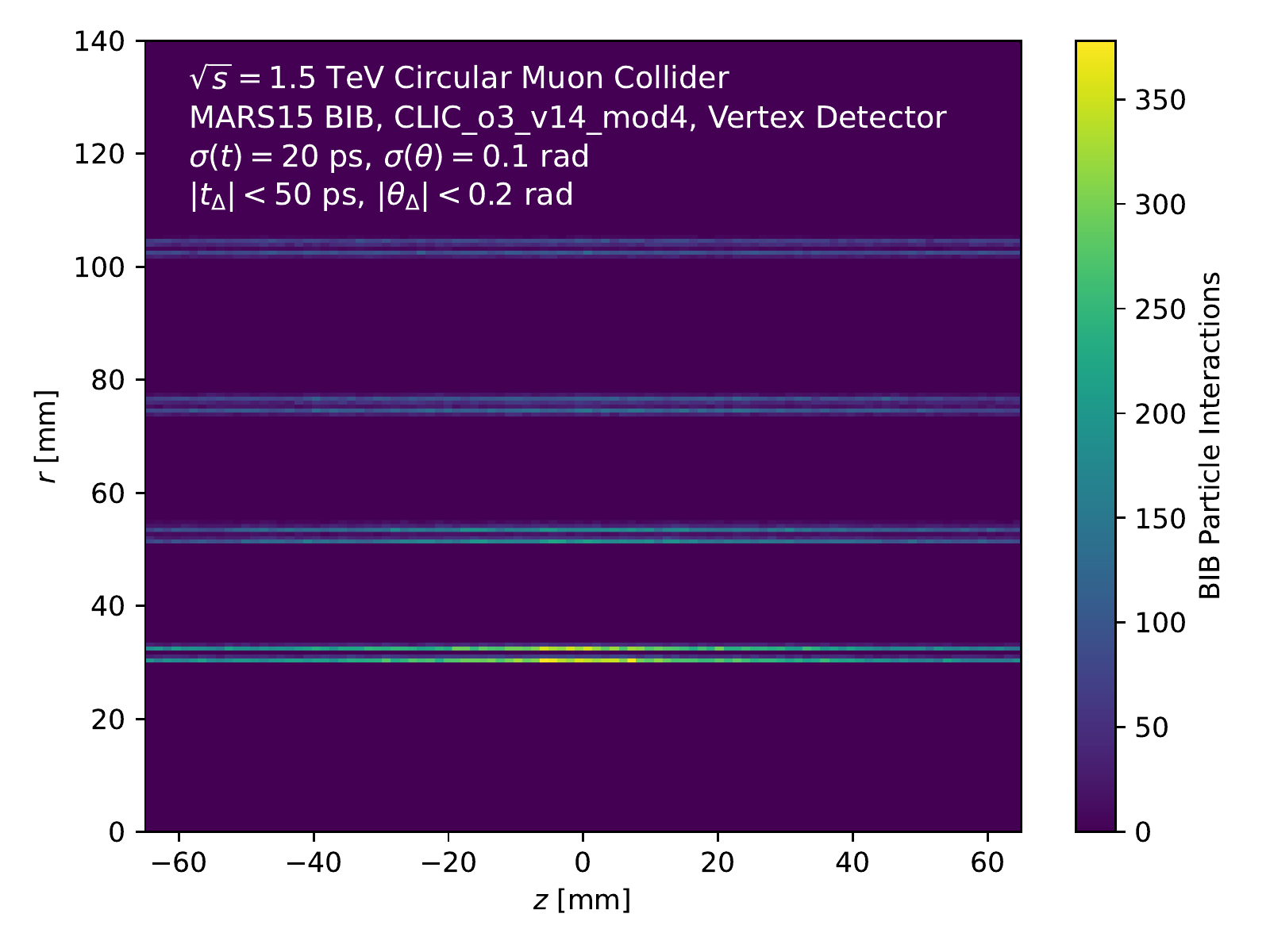}
    \caption{The spatial distribution of vertex detector sim hits is shown before [Left] and after [Right] a requirement of $|t_\Delta|<50$~ps and $|\theta_\Delta|<0.2$~rad is applied for $(\sigma(t),\sigma(\theta))=(20\mbox{ ps},0.1\mbox{ rad})$. On the left, BIB contributions on the first tracking layer from the forward regions closest to the nozzles is apparent. After placing these requirements to reject BIB, the only remaining components are those that mirror the distributions of collision products, emerging approximately from the IP. The spatial distribution is shown in transverse radius $r$ and longitudinal distance $z$.}
    \label{fig:spatialDist}
\end{figure}

Performance of the two-dimensional discriminator for rejecting BIB in the tracker can be found in Figure~\ref{fig:ROC} for various measurement resolutions as defined by a gaussian smearing of the measured $t$ and $\theta$ values. Since there are two rectangular cuts, a ROC \emph{ribbon}, comprising a collection of individual ROC curves, can be calculated. A single point along one of these ribbons corresponds to the performance of a single set of cuts and resolution values. For a set of resolutions $(\sigma(t),\sigma(\theta))=(20\mbox{ ps},0.1\mbox{ rad})$, collision particle interactions can be selected with an efficiency of $99\%$ while rejecting $81.5\%$ of the fiducial contribution from BIB. Poorer resolutions at $(\sigma(t),\sigma(\theta))=(100\mbox{ ps},0.1\mbox{ rad})$ result in only $60.5\%$ of the fiducial BIB contribution being rejected.  The spatial distribution of vertex detector sim hits is shown in Figure~\ref{fig:spatialDist} before and after the application of a requirement that $|t_\Delta|<50$~ps and $|\theta_\Delta|<0.2$~rad for $(\sigma(t),\sigma(\theta))=(20\mbox{ ps},0.1\mbox{ rad})$.

This performance is highly dependent on the geometry of the vertex detector and the size of the beamspot. Because the vertex detector barrel is close to the beamline and extends only to 6.5 cm in $z$, the 1 cm beamspot spread means that a large range of angles are consistent with collision products. A study performed with a 1 mm beamspot showed that the BIB could be reduced an additional order of magnitude, so long as a detector angular resolution of 0.01 rad could be achieved to take advantage of it. For detectors at larger distances from the IP, this dependence would be smaller, making this strategy interesting for larger radius trackers and the calorimeter. 

\section{Angular Discrimination in Calorimeters}
\label{sec:calo}

\begin{figure}[htb]
    \centering
    \begin{tikzpicture}
        \node[anchor=south west](image) at (0,0) {
            \frame{\includegraphics[trim=10 150 100 150,clip,width=0.65\textwidth]{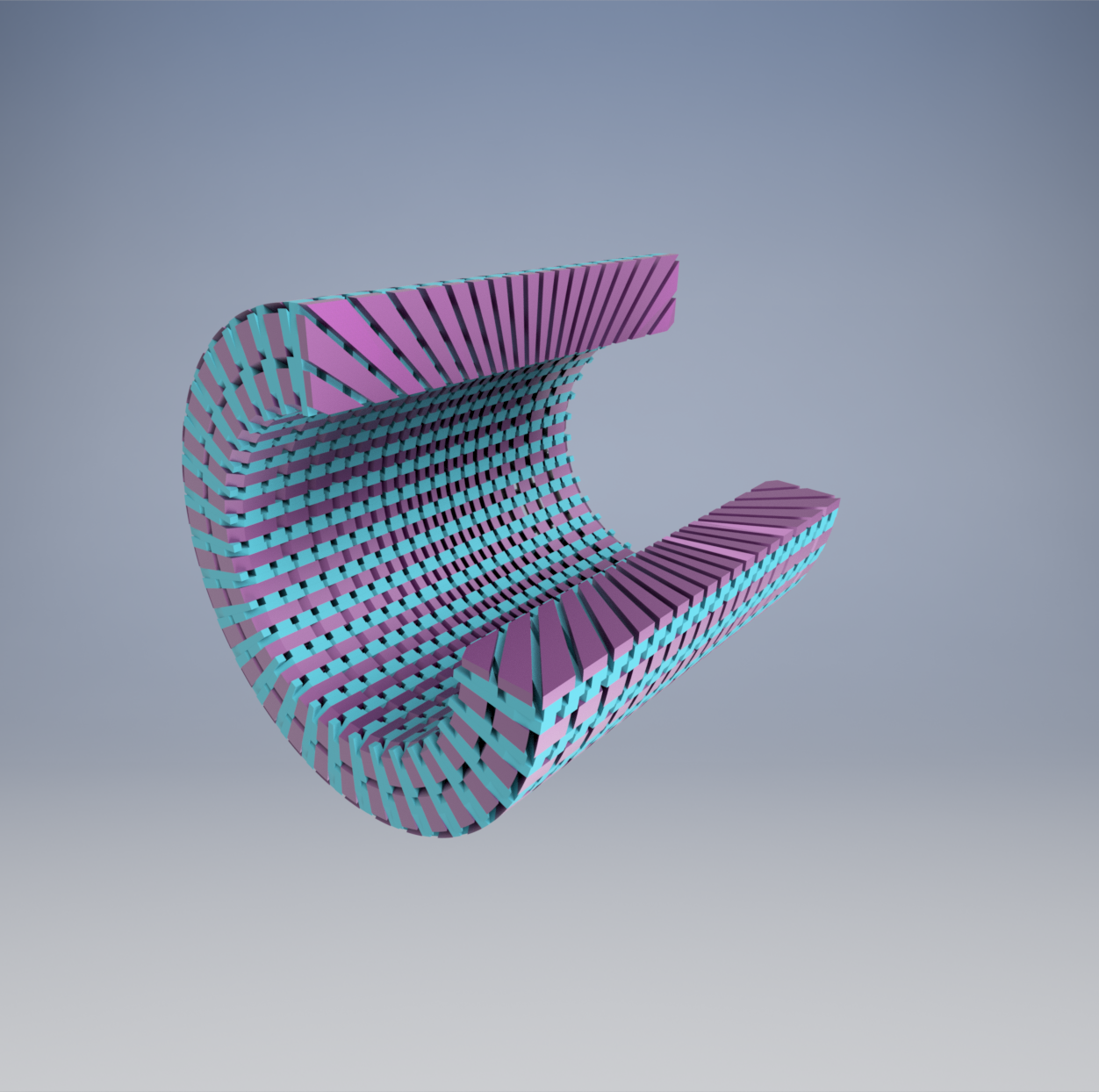}}\vspace{0.2em}
        };
        \begin{scope}[
            x={($0.1*(image.south east)$)},
            y={($0.1*(image.north west)$)}
            ]
            \node[align=right] (A) at (7.0,1) {\textbf{Conceptual Design} \\ \textbf{for a Triscopic Calorimeter}};
        \end{scope}
    \end{tikzpicture}
    \begin{tikzpicture}[font=\tiny]
        \centering
        \node[anchor=north east](image2){\frame{\includegraphics[trim=80 150 100 150 ,clip,width=0.4\textwidth]{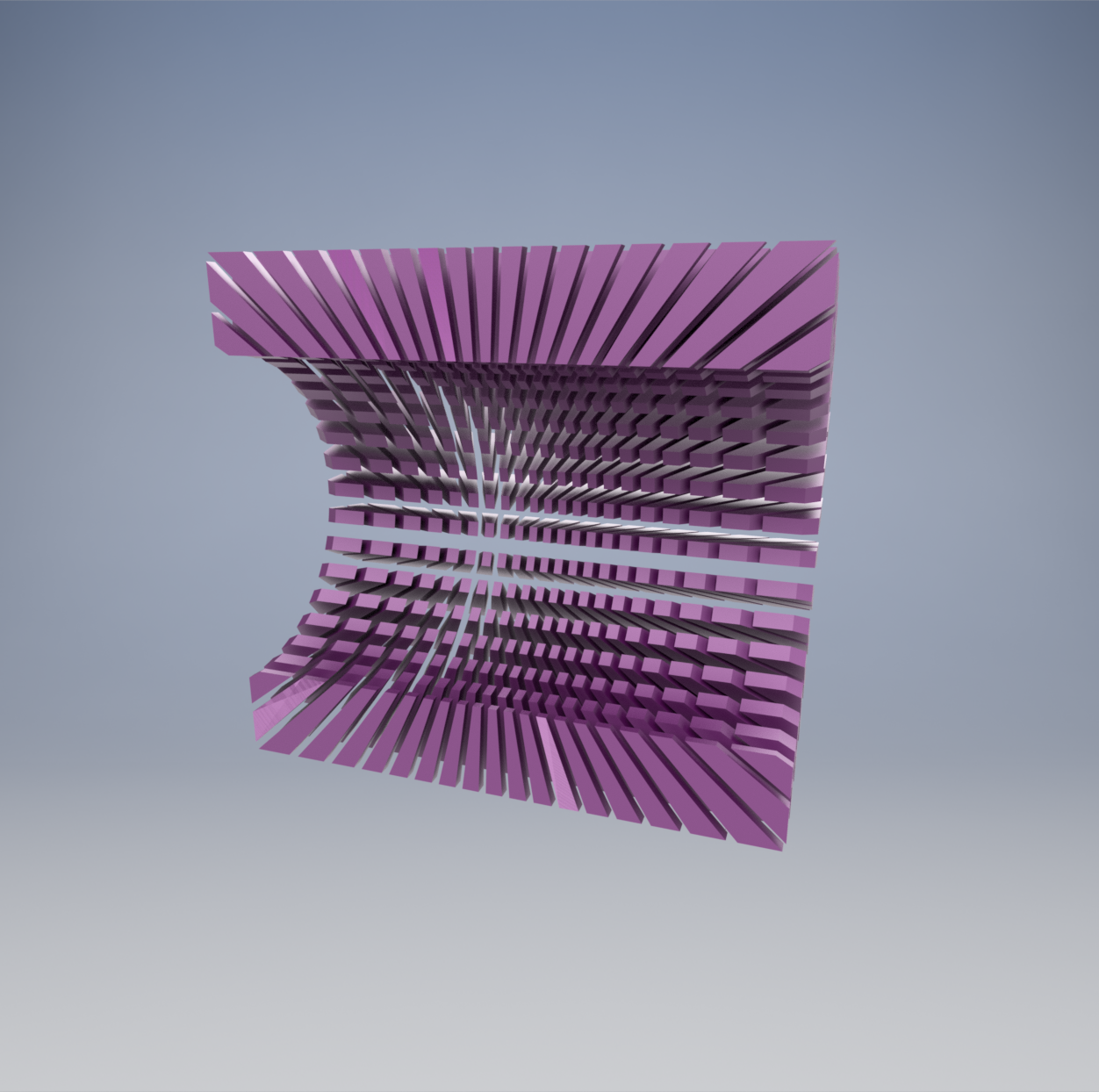}}
        };
        \begin{scope}[
            x={($0.1*(image2.south east)$)},
            y={($0.1*(image2.north west)$)}
        ]
            \node[align=left] (B) at (9,6.7) {\textbf{Conceptual Design}\\ \textbf{for a Triscopic Calorimeter}};
        \end{scope}
    \end{tikzpicture}
    \begin{tikzpicture}[font=\tiny]
        \centering
        \node[anchor=north west](image3){\frame{\includegraphics[trim=80 150 100 150 ,clip,width=0.4\textwidth]{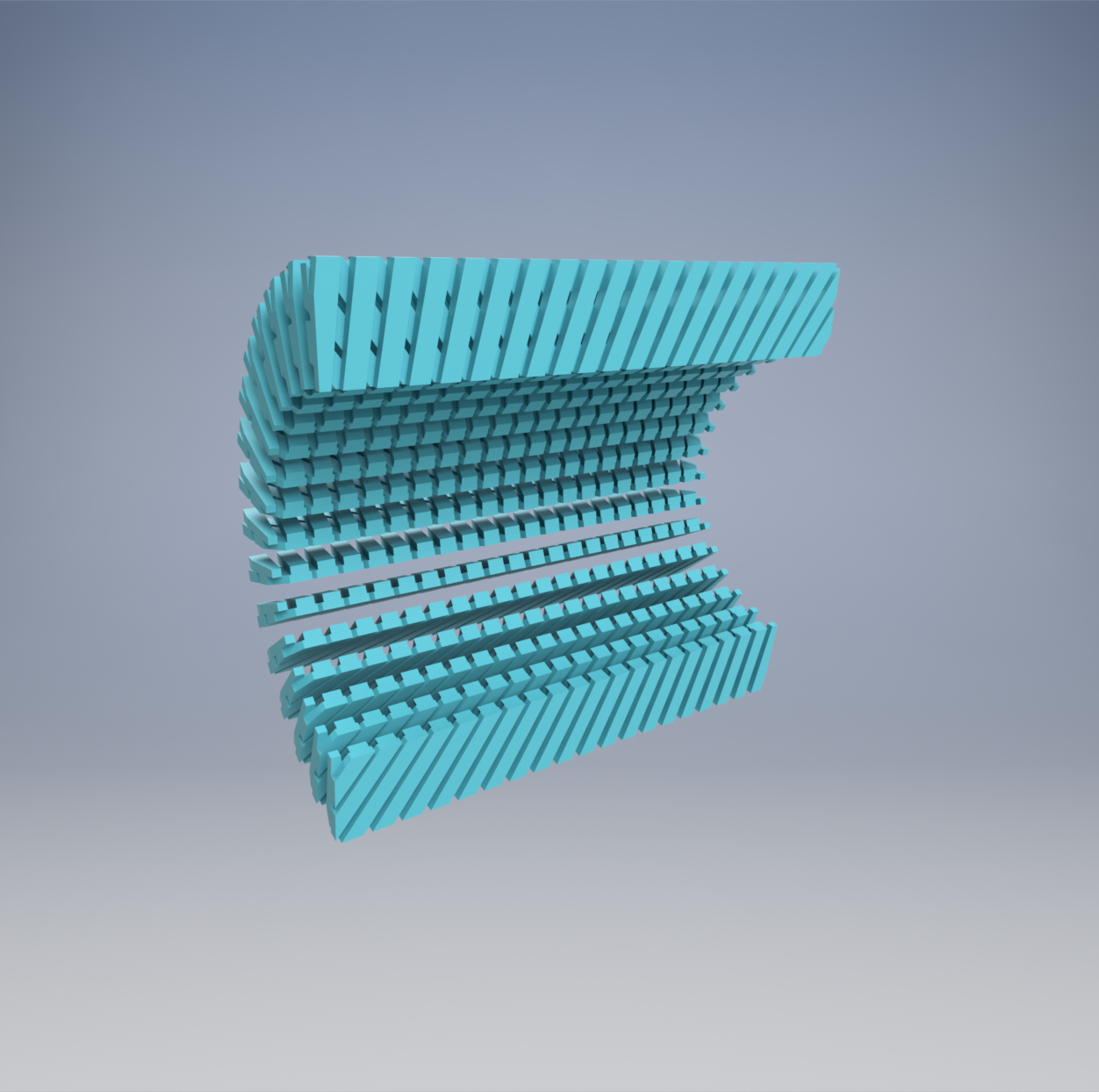}}
        };
        \begin{scope}[
            x={($0.1*(image3.north east)$)},
            y={($0.1*(image3.south west)$)}
        ]
            \node[align=right] (C) at (6.7,9) {\textbf{Conceptual Design}\\ \textbf{for a Triscopic Calorimeter}};
        \end{scope}
    \end{tikzpicture}
    \caption{[Top] A three-quarters cut-away view of the proposed triscopic calorimeter is shown. IP slices are shown in purple while the BIB slices are shown in blue. [Bottom Left] A half view of the IP slices is shown. Elements are aligned such that they are projective from the IP. [Bottom Right] The BIB slices are shown, inter-weaved such that half of the layers are angled to be consistent with BIB from each beam.
    }
    \label{fig:calo1}
\end{figure}

In calorimeter systems, precise timing information may be difficult to achieve. Angular information can be used for the rejection of BIB, but the geometry of the detector elements will greatly affect this discrimination power. At the LHC, calorimeter elements are generally designed with the IP in mind, prioritizing energy deposition in tower volumes that are projective from the IP. For example, the ATLAS LAr calorimeter comprises square tower units in sampling layer 2 that have a volume of roughly $\Delta\eta\approx\Delta\phi\approx0.025$ with a radial length of roughly $450$~mm. Integrating energy deposition over this entire volume would be very sensitive to the high-BIB environment of a muon collider.

\begin{figure}[tb]
    \centering
    \begin{tikzpicture}
        \centering
        \node[anchor=south west] (image4){
        \hspace{0.02\textwidth}\frame{\includegraphics[trim=50 50 50 0,clip,width=0.75\textwidth]{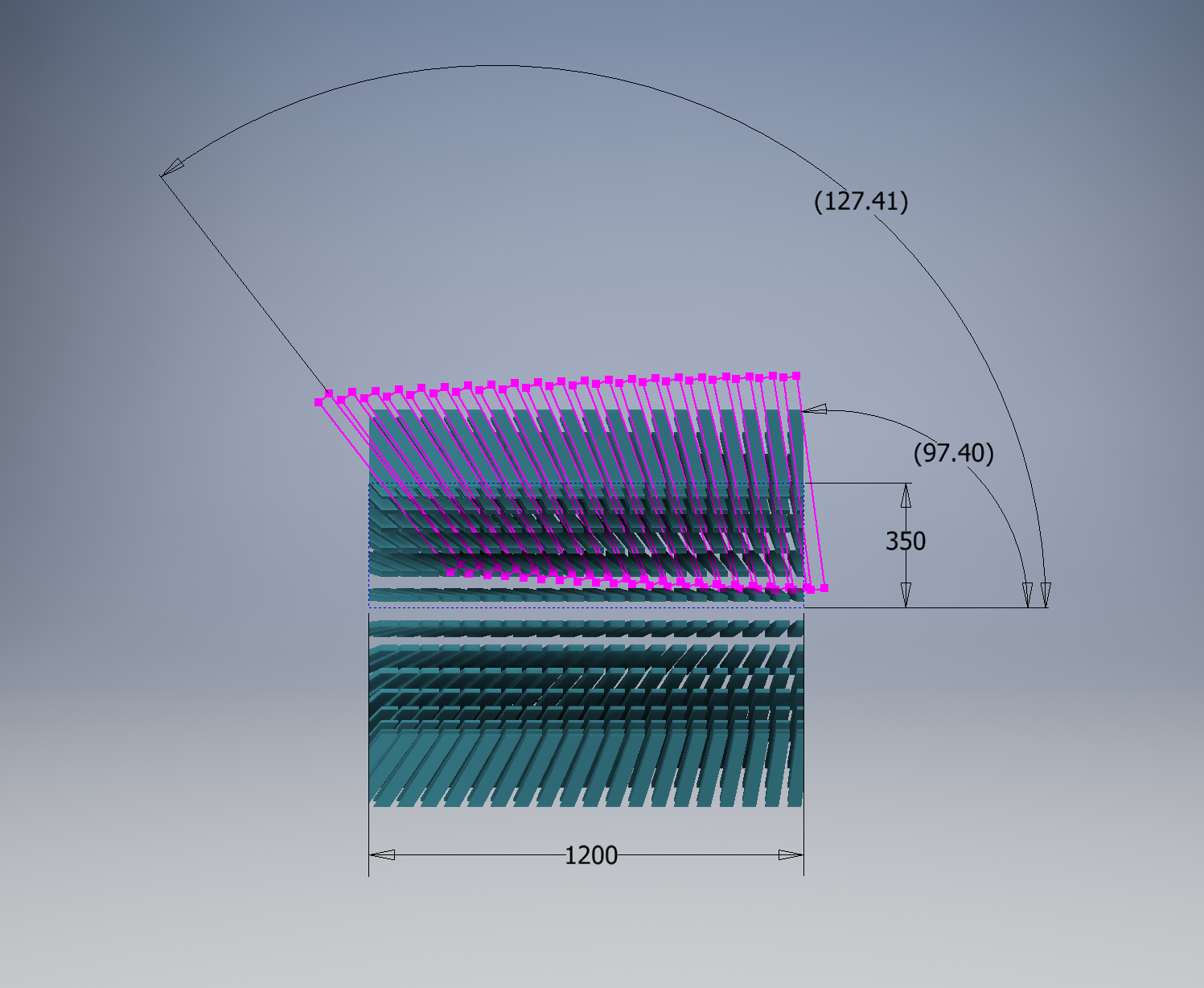}}};
        
        \begin{scope}[
            x={($0.1*(image4.south east)$)},
            y={($0.1*(image4.north west)$)}
            ]
            \node[align=left] (D) at (1.95,1) {
\footnotesize\textbf{Conceptual }\\\footnotesize\textbf{Design for a }\\ \footnotesize\textbf{Triscopic Calorimeter}};
        \end{scope}
    \end{tikzpicture}
    \caption{An example BIB slice is shown, targeting BIB due to the beam entering the detector from the right. The optimal angle to accept BIB varies across the longitudinal length of the detector from $\theta=97.40^\circ$ to $127.41^\circ$ as pictured. The opposing BIB slice mirrors this structure. The linear dimensions are arbitrary and for drawing purposes only.
    }
    \label{fig:calo2}
\end{figure}

Using the distributions shown in Figure~\ref{fig:angleDists}, assuming the angles of incidence are representative of what the calorimeter would see, we propose a toy calorimeter geometry designed to give offline analysis as much information as possible to distinguish BIB from the collision products. This geometry contains $\phi$-segmentation of three alternating types. A particular slice of the calorimeter in $\phi$ will be a) an ``IP'' slice designed with IP-projective towers, b) a ``BIB-1'' slice designed for particles consistent with BIB from Beam 1, or c) a ``BIB-2'' slice corresponding to BIB from Beam 2. The detector $\theta-\phi$ segmentation can be fine enough such that a typical hadronic jet deposits energy in all of these types of layers, giving a \emph{triscopic} calorimeter measurement. The triscopic angles vary across the detector to account for the changing incidence angles along the beam-line. An example geometry demonstrating this structure can be seen in Figure~\ref{fig:calo1}. The angular structure of the BIB layers, as derived from Figure~\ref{fig:angleDists}, can be seen in Figure~\ref{fig:calo2}.

A jet that has been reconstructed in the calorimeter can be characterized by the particular pattern of triscopic calorimeter signals that form the jet. For example, defining a width measure in the longitudinal location $z$ as

\begin{align}
    \sigma_z = \sum_i \frac{E_i}{E_{\rm{total}}} \Delta z_i,
\end{align}

\noindent where the sum runs over the calorimeter measurements that compose the jet, and $\Delta z_i$ is the longitudinal distance between the measurement and the longitudinal location of the center of the jet. 

Measuring $\sigma_z$ exclusively in each type of $\phi$-layer, allows for comparisons between layers. For each jet, we can construct a quantity such as

\begin{align}
    R_{\rm BIB} = \frac{\min{\sigma_{z,\rm BIB}} }{\sigma_{z,\rm IP} } \cdot \frac{\min \sigma_{z,\rm BIB} }{\max \sigma_{z,\rm BIB} } = \frac{\min \sigma_{z,\rm BIB}^2 }{\sigma_{z,\rm IP}\ \max\sigma_{z,\rm BIB} },
\end{align}

\noindent where $\min$ and $\max\sigma_{z,\rm BIB}$ represent the smaller and larger of the two values $\sigma_{z,\rm BIB-1}$ and $\sigma_{z,\rm BIB-2}$, respectively. This variable can discriminate jets originating from BIB and those originating from the collision -- BIB-induced jets from either beam will result in a low value of $R_{\rm BIB}$, while collision jets will have a relatively large value. In general, the triscopic calorimetry is designed to give as much information to offline analysis as possible, prioritizing the identification and rejection of BIB. For triggered readout systems, this information can also easily be included into the calorimeter trigger systems to reduce the rate of the jet triggers due to BIB.

\section{Conclusions and Future Works}

In the vertex detector, the combination of timing and angular information can be used to build a discriminator that strongly rejects BIB. Even after a tight fiducial timing window, nearly an order of magnitude of additional BIB rejection can be achieved with realistic timing detectors and angular measurements from paired layers. For a detector with 20 ps timing resolution and 0.1 rad angular resolution,  90\% BIB reduction can be achieved while retaining 95\% of signal hits. This performance is highly dependent on the size of the beamspot; the reduction of the beamspot by a factor of 10 increased rejection power by approximately an order of magnitude.

In the calorimeter, this kind of precision is less feasible, but dedicated detector geometries can be explored to help lessen the impact of BIB. However, in this comparatively high-$r$ and high-$z$ detector, the impact of the beamspot size is much smaller. More studies are needed to determine the quantitative impact of these designs.

These rejection techniques are all predicated on the assumption that signal processes originate at the IP, so tight cuts necessarily reduce sensitivity to the decays of long-lived particles. Further studies are required to fully explore this trade-off. 

\acknowledgments
We thank the University of Wisconsin-Madison experimental HEP group for useful discussions and computing, with particular thanks to S. Dasu and C. Vuosalo. We also thank the larger community of Muon Collider enthusiasts for useful feedback, especially the International Muon Collider Collaboration, S. Jindariani, and S. Pagan Griso. This work was supported by U.S. Department of Energy, Office of Science, Office of Basic Energy Sciences Energy Frontier Research Centers program under Award Number DE-SC0020267 (T.H., D.A., L.C.), and Award Number DE-SC0007881 (L.L.).

\printbibliography

\end{document}